\documentclass[prl,reprint,superscriptaddress,showpacs]{revtex4-1}
\usepackage{graphicx}
\usepackage[utf8]{inputenc}
\usepackage{amsmath}
\usepackage{dcolumn}

\begin{document}

\title{Shot Noise of a Temperature-Biased Tunnel Junction}

\author{Samuel Larocque}
\author{Edouard Pinsolle}
\author{Christian Lupien}
\author{Bertrand Reulet}
\affiliation{Institut Quantique, D\'{e}partement de Physique, Universit\'{e} de Sherbrooke, Sherbrooke, Qu\'{e}bec J1K 2R1, Canada.}
\date{\today}

\begin{abstract}
	We report the measurement of the current noise of a tunnel junction driven out-of-equilibrium by a temperature and/or voltage difference, i.e. the charge noise of heat and/or electrical current. This is achieved by a careful control of electron temperature below 1 K at the nanoscale, and a sensitive measurement of noise with wide bandwidth, from 0.1 to 1 GHz. An excellent agreement between experiment and theory with no fitting parameter is obtained. In particular, we find that the current noise of the junction of resistance $R$ when one electrode is at temperature $T$ and the other one at zero temperature is given by $S=2\,\mathrm{ln}2k_B T/R$.   
\end{abstract}

\maketitle
Caloritronic in small systems has been of high interest in recent years as new ways to manipulate electronic heat currents at the nanoscale have been developed \cite{giazotto_opportunities_2006}. The ability to make mesoscopic systems with well controlled temperature \cite{white_status_1996,molenkamp_peltier_1992} enables the study of heat transport \cite{martinez-perez_rectification_2015,hwang_phase-coherent_2018,meschke_single-mode_2006,ronzani_tunable_2018} and quantum thermodynamics \cite{molenkamp_peltier_1992,jezouin_quantum_2013,chiatti_quantum_2006}, both of fundamental interest. A tunnel barrier between two metallic electrodes forms the basic unit in the study of non-equilibrium physics and fluctuations \cite{fevrier_tunneling_2018}. This system has been put to great use in the understanding of electronic transport whether at equilibrium where fluctuations are used as a thermometer \cite{spietz_primary_2003, pinsolle_direct_2016} or in the presence of voltage bias where information on charge carrier can be accessed \cite{saminadayar_observation_1997,kapfer_josephson_2019,de-picciotto_direct_1998}. Very recently, a study has been carried out to study the effect of small temperature gradient in such a system \cite{lumbroso_electronic_2018}.

In this letter we present calculations and measurements of the electrical noise in a metallic tunnel junction in the presence of arbitrary thermal and voltage gradients. With the ability to work at very low temperature, this experiment is not limited to small temperature differences and mixes heat and charge transport thus extending greatly the work started in \cite{lumbroso_electronic_2018}.

\emph{Theory.} Using the scattering matrix formalism \cite{blanter_shot_2000}, the current noise spectral density $S_{2}$ of a tunnel junction of transmission $D\ll1$ (supposed energy independent for the sake of simplicity) is related to the energy distribution functions in the left $f_L$ and right $f_R$ reservoirs by:
\begin{equation}
\begin{aligned}
	& S_{2}=\frac{e^2D}{\pi \hbar} \int dE [f_L(E) \left( 1 - f_L(E)  \right) \\
	& + f_R(E) \left( 1 - f_R(E) \right)+\left( f_L(E) - f_R(E) \right)^2]
\end{aligned}
\label{eq:Fermi}
\end{equation}

\noindent with $e$ the electron charge and $\hbar$ the Planck constant. In the presence of a voltage bias $V$ but no thermal gradient on the junction, $f_L$ and $f_R$ are two Fermi-Dirac distributions at temperature $T$ with chemical potentials shifted by $eV$, and Eq.(\ref{eq:Fermi}) leads to the well known formula for the shot noise at low frequency:

\begin{equation}
\begin{aligned}
	S_{2}=eI\coth{\frac{eV}{2k_BT}}
\end{aligned}
\label{eq:ShotN}
\end{equation}

\noindent where $k_B$ is the Boltzmann constant and $I$ the electrical current flowing through the junction. However, to our knowledge, no analytical expression has been derived in the case of a thermal gradient applied to the junction. We have obtained such an expression in two limits. First, when the temperatures of the hot electrode $T_{\text{Hot}}$ and the cold one $T_{\text{Cold}}$ are very close, the noise generated by the junction can be approximated by: 
\begin{equation}
	S_{2}=\frac{2 k_B}{R_j} \left(  \frac{T_{\text{Cold}} + T_{\text{Hot}}}{2} \right)
\label{eq:TL=TR}
\end{equation}

\noindent with $R_j$ the electrical resistance of the tunnel junction. This limit, which has been studied in detail in \cite{lumbroso_electronic_2018}, corresponds to the Johnson-Nyquist noise \cite{nyquist_thermal_1928,johnson_thermal_1928} of the junction at a temperature $(T_{\text{Cold}}+T_{\text{Hot}})/2$.
The other interesting limit is when the cold electrode is at zero temperature, $T_{\text{Cold}}=0$, for which we find: 
\begin{equation}
	S_{2} = \frac{2\; \mathrm{ln}\,2\;k_B}{R_j} T_{\text{Hot}}
\label{eq:TLggTR}
\end{equation}
In the case of arbitrary temperature/voltage differences, we have performed numerical calculations of $S_2$ to compare with our experiment, as shown below.

The $\mathrm{ln}\,2$ factor in Eq.(\ref{eq:TLggTR}) comes from the integral of the Fermi-Dirac function over positive energies, which counts the number of excited electrons in the electrode at temperature $T_{\text{Hot}}$, and that of $1-f$ over negative energies, which counts the number of holes, equal to that of electrons:
\begin{equation}
	N_e=\int_0^{+\infty}n(E)f(E)dE=\mathrm{ln}\,2k_BT_{\text{Hot}}n(0)
\end{equation}
with $n(E)$ the density of state (supposed energy independent) and $n(0)$ its value at Fermi energy. The fact that the zero-frequency noise measures the total number of excitations has been already discussed in the context of minimal excitation pulses \cite{dubois_minimal-excitation_2013}. The $\mathrm{ln}\,2$ factor also recalls the Landauer limit for the erasure process of a bit of information \cite{landauer_irreversibility_1961}, and there might be deeper links between Eq.(\ref{eq:TLggTR}) and information theory. To be more precise, electrons crossing the junction in one direction or the other generate respectively positive and negative current pulses, to which one can associate a bit of information, 0 or 1. The noise measures the number of bits per second emitted by the junction. The electrons that cross the barrier leave a hole behind, but thermalisation in the electrodes of the junction ensures that this hole is filled to keep the energy distribution constant. Thus the information associated to the crossing of electrons, i.e. to the emission of a bit of information, is erased in the reservoirs.


\begin{figure}[t!]
\center
	\includegraphics[width=1\columnwidth]{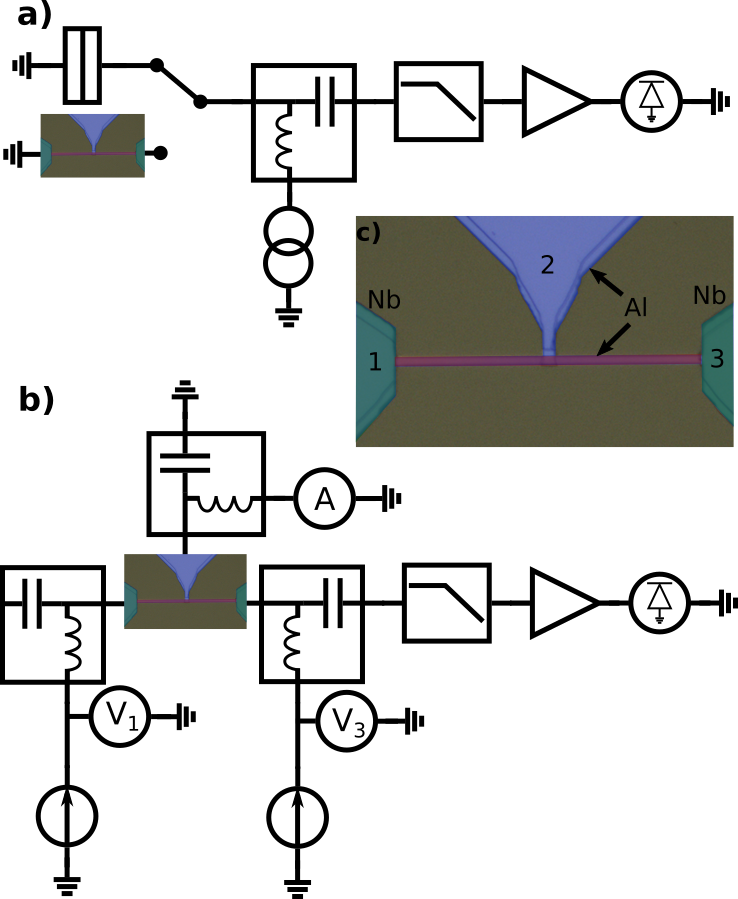}

	\caption{a) Experimental setup used to calibrate the noise generated by the wire as a function of applied dc current. A cryogenic switch allows to measure either the wire (between contacts 1 and 3) or a reference tunnel junction. Contact 2 of the sample is left unconnected. b) Experimental setup for the measurement of the tunnel junction under voltage and temperature bias. Contacts 1 and 3 are voltage biased independently while the noise generated by the junction is measured. c) Photograph of the sample. The wire is between the Nb contacts 1 and 3; the junction is between the wire and the Al contact 2.}

\label{fig:setup}
\end{figure}

\emph{Principle of the experiment.} In order to create a temperature difference between the two electrodes of the junction, we attach the junction on one side to a diffusive wire and on the other side to a large electrode. The latter allows efficient electron thermalisation, so that the electron temperature in that electrode is that of the phonon bath: $T_{\text{Cold}}=T_{\text{ph}}$. By imposing a dc current in the wire, but not in the junction, we can control the amount of Joule heating and thus the electron temperature in the wire. The temperature profile in the wire and hence the electron temperature at the junction $T_{\text{Hot}}$, strongly depends on the cooling mechanisms at play. At low temperature, these are predominantly electron-phonon interaction and hot electrons diffusing out of the sample. By contacting the wire to superconducting electrodes, we block diffusion cooling. Since electron-phonon interaction is uniform along the wire, the electron temperature is also uniform, given by:
\begin{equation}
	T_w=\left( \frac{R_w I_w^2}{\Sigma \Omega}+T_{\text{ph}}^n\right)^{1/n}
\label{eq:eph}
\end{equation}

\noindent with $I_w$ the current inside the wire, $R_w$ its electrical resistance, $\Omega$ its volume, $\Sigma$ the electron-phonon coupling constant and $n$ a power law that typically varies between 4 and 6 depending on the material.


In a first experiment we measured the noise \emph{of the wire} as a function of the voltage difference between its contacts $V_w$. This permits to link the Joule power dissipated in the wire to its electron temperature and hence to know the temperature of the hot electrode of the junction $T_{\text{Hot}}$. Then, in another experiment, we measured the noise \emph{of the junction}  as a function of both voltage and temperature gradients across the junction.\\

\emph{Sample fabrication}. A photograph of the sample is given in Fig. \ref{fig:setup}(c). It consists of a $2\times 3\,\mu$m aluminum tunnel junction of resistance $R_{j}=1300\,\Omega$ between a large contact (\#2 in Fig. \ref{fig:setup}(c)) and a thin wire, both made of Al. The contact is large ($300\,\mu$m$\times300\,\mu$m) and thick ($200\,\text{nm}$) in order to stay at the cryostat temperature for all bias \cite{henny_1/3-shot-noise_1999}. The wire, of resistance $R_{w}=165\,\Omega$, is $100\,\mu$m long, $2\,\mu$m wide, $25\,\text{nm}$ thick, with the junction in its middle.  It is contacted to two niobium reservoirs (\#1 and \# 3 in Fig. \ref{fig:setup}(c)) identical to the Al contact of the junction. 
The sample has been made in two steps. First the wire and the junction are fabricated by photo-lithography followed by a shadow evaporation of Al through a Dolan bridge \cite{dolan_offset_1977}. Then the Nb reservoirs are added to the wire, by e-beam evaporation after removing the native Al oxide by ion milling to ensure a low resistance contact. The sample is placed on the 7 mK stage of a dilution refrigerator. A magnetic field of 100 mT is applied to the sample to turn the Al normal while keeping the Nb superconducting. \\

\begin{figure}[t!]
\center
	\includegraphics[width=1\columnwidth]{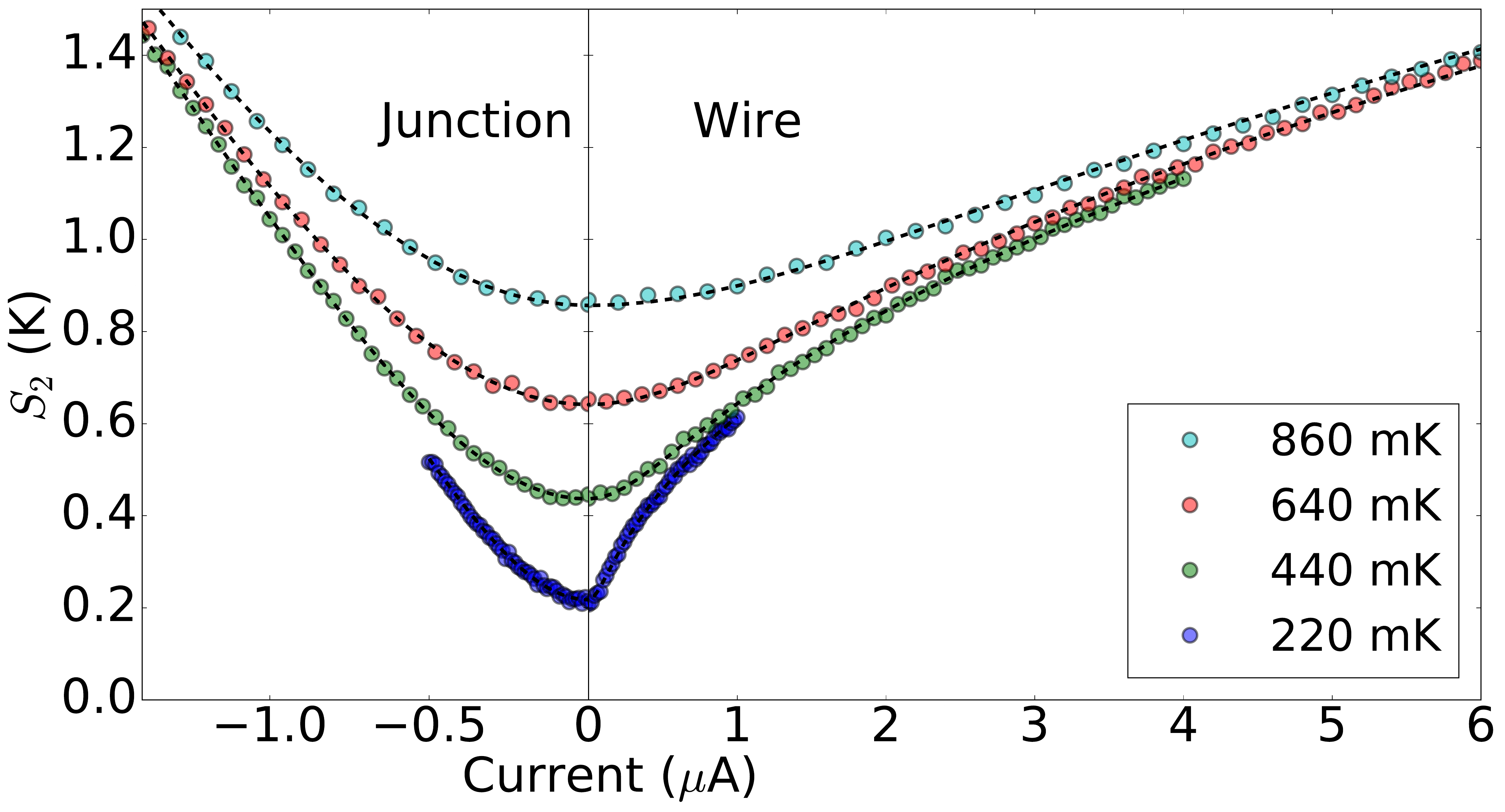}

	\caption{Noise temperature of the wire (right part) and of the reference tunnel junction (left part) as a function of applied dc current. Symbols of different colors correspond to experimental data taken at different phonon temperatures. Dashed lines correspond to theoretical predictions of Eq. (\ref{eq:ShotN}) for the junction and Eq.(\ref{eq:eph}) for the wire.}

\label{fig:fil}
\end{figure}

\emph{Experiment \#1: calibration.} The experimental setup for the first experiment, i.e. the calibration of the electron temperature in the wire, is presented in Fig \ref{fig:setup}(a). In this setup, one end of the wire (\# 1) is connected to ground and the other end (\# 3) to the measurement circuit. The contact of the junction (\# 2) is left unconnected.
 
The wire is current biased through the dc-port of a bias-tee. The ac port of the bias-tee is connected to a cryogenic microwave amplifier placed at 3 K and the noise in the range of 40 MHz-1 GHz is measured by a power detector. A cryogenic microwave switch enables to change in-situ from the wire to a well known tunnel junction of resistance $R_{ref}=177\,\Omega$ (i.e., very close to the one of the wire) used as a reference sample enabling the calibration of the system (gain of the amplification chain and noise of the amplifier). In the following we will express the noise $S_2$ of all devices in terms of an equivalent noise temperature $T_{\text{Noise}}$ given by $S_2=2k_BT_{\text{Noise}}/R$ with $R$ the resistance of the device, wire or junction. The noise temperature of the heated wire is simply given by its electron temperature: $T_{\text{Noise}}=T_w$.
 
\begin{figure}[t!]
\center
	\includegraphics[width=1\columnwidth]{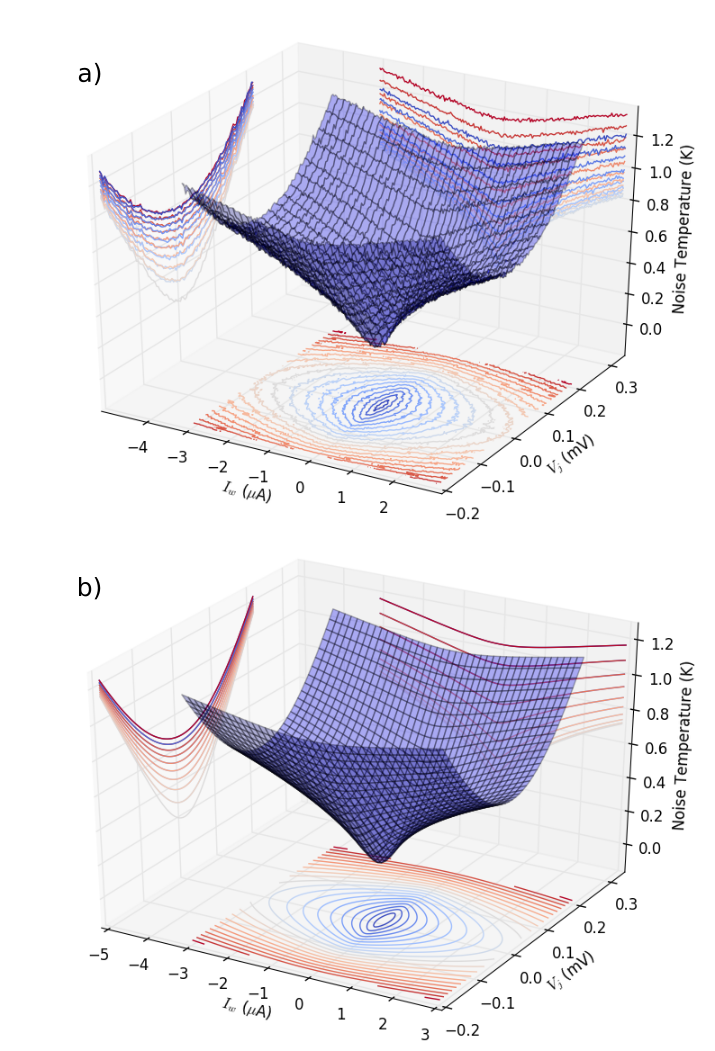}
	\caption{Noise temperature of the tunnel junction as a function of voltage bias $V_j$ on the junction and heating current $I_w$ in the wire for a phonon temperature $T_{ph}=200$ mK. The top figure corresponds to experimental data and the bottom one to numerical evaluations of Eq.(\ref{eq:Fermi}).}
\label{fig:3D}
\end{figure}
At low frequency, the noise measured by our setup is given by:
\begin{equation}
S_M=G_{\text{eff}}(S_A+(1-\Gamma^2)S_2)
\label{eq:2}
\end{equation}
with $G_{\text{eff}}$ the effective gain of the setup, $S_A$ the noise generated by the amplifier and $\Gamma=(R-R_0)/(R+R_0)$ the reflection coefficient of the sample (having a reference junction of resistance close to that of the wire avoids systematic errors due to imprecisions on $\Gamma$). The left part of Fig. \ref{fig:fil} shows the noise temperature of the reference junction as a function of the applied current for various phonon temperatures from 50 mK to 850 mK. The measured noise of the junction is very well fitted by Eqs. (\ref{eq:2}) and (\ref{eq:ShotN}), thus providing the effective gain $G_{\text{eff}}\simeq10^{9}$ of the entire setup and the noise temperature of the amplifier, $T_A=3.2$ K. The knowledge of these parameters enables the calibration of the noise generated by the wire, i.e. provides the link between the measured noise and the electron temperature in the wire. The electron temperature of the wire vs. dc current is shown in the right part of Fig. \ref{fig:fil}. It is accurately fitted by Eq. (\ref{eq:eph}) using $\Sigma=1.3\pm 0.1 \times 10^{-9}\,\mathrm{W K}^{-5} \mu \mathrm{m}^{-3}$ and $n=4.60 \pm 0.01$. These values are usual for Al thin films \cite{pinsolle_direct_2016,giazotto_opportunities_2006}. This first measurement allows to control the temperature of the hot electrode of the tunnel junction $T_{\text{Hot}}$, by adjusting the dc current in the wire $I_w$.\\


\emph{Experiment \#2: noise of the temperature biased junction.} The setup used for the second measurement is shown in Fig. \ref{fig:setup}(b). Here the three contacts of the sample are connected to three bias tees, to allow independent control of the current in the wire and the junction. It is biased through dc ports of bias tees on contacts 1 and 3, the voltage of which is measured by two voltmeters connected to the bias tees using independent wiring. In principle, if we apply opposite voltage on contacts 1 and 3, i.e. $V_3=-V_1$, there should be no electrical current in the junction, provided the junction sits exactly in the middle of the wire and that there are no thermoelectric voltages in the setup. In order to be sure that we can achieve a pure heat current with no electrical current in the junction, we measure the latter  by connecting the dc port of the bias tee connected to the junction (contact 2) to an ammeter, thus measuring directly the dc electrical current through the junction. In contrast, if we apply equal voltages on contacts 1 and 3, $V_3=V_1$, we achieve a situation where there is very little heating of the wire with a finite dc voltage $V_j$ across the junction. The ac port of the bias tee of contact 3 is connected to the cryogenic amplifier followed by a power meter as in the previous setup, while the ac port on contact 2 is connected to ground and the one of contact 1 is left open. Thus this setup measures the noise of the junction in series with two half wires in parallel. This corresponds to a total resistance of $R_{\text{tot}}=R_j+R_w/4$ and a total noise temperature $T_{\text{tot}}$ given by:
\begin{equation}
	T_{\text{tot}}=\frac{R_j}{R_{\text{tot}}}T_j+\frac{R_w}{R_{\text{tot}}}T_w
\label{eq:total_noise}
\end{equation}
with $T_j$ the noise temperature of the junction and $T_w=T_{\text{Hot}}$ that of the wire. The choice of a junction of resistance $R_j$ much higher than that of the wire $R_w$ has several purposes: i) it makes the contribution of the wire to the measured noise negligible in Eq.(\ref{eq:total_noise}): $T_{\text{tot}}\simeq T_{j}$. The noise measured with this setup is thus simply that of the junction. ii) It reduces the heating of the wire while biasing the junction. This is important for $eV_j\sim k_BT_{\text{Cold}}$ (at high bias, the noise of the junction is independent of the temperature of the contacts). At $T=220$ mK and a bias $V_j=k_BT/e\sim20\,\mu$V, a current of $I_w=V_j/(2R_{\text{tot}})=7.5\,\mathrm{nA}$ is flowing in each half of the wire, which leads to a negligible temperature increase of $T_{\text{Hot}}$, see Fig. \ref{fig:fil} right. iii) It avoids cooling of the electrons in the wire by conduction through the junction.

By biasing the junction with the least current flowing in the wire (nominally $V_1=V_3=V_j$) we observe the usual shot noise of a voltage biased tunnel junction given by Eq. \eqref{eq:ShotN}. This provides a calibration of the setup, as in the first experiment. Once the setup is calibrated, a current $I_w$ is applied on the wire while keeping no current in the junction (nominally $V_1=-V_3$). In this situation, the tunnel junction generates noise that is only due to the temperature difference, i.e. with a pure heat current flowing through it, as desired. Composite regimes are achieved by adjusting $V_1$ and $V_3$. The results of such measurements are shown in Fig. \ref{fig:3D}(a) where the noise temperature of the junction is plotted as a function of both voltage bias $V_j\simeq(V_1+V_3)/2$ and current in the wire $I_w\simeq(V_1-V_3)/R_w$. In Fig. \ref{fig:3D}(b) we show corresponding  numerical simulations with no adjustable parameters. There is a very good agreement between theory and experiment.

\begin{figure}[ht!]
\center
	\includegraphics[width=1\columnwidth]{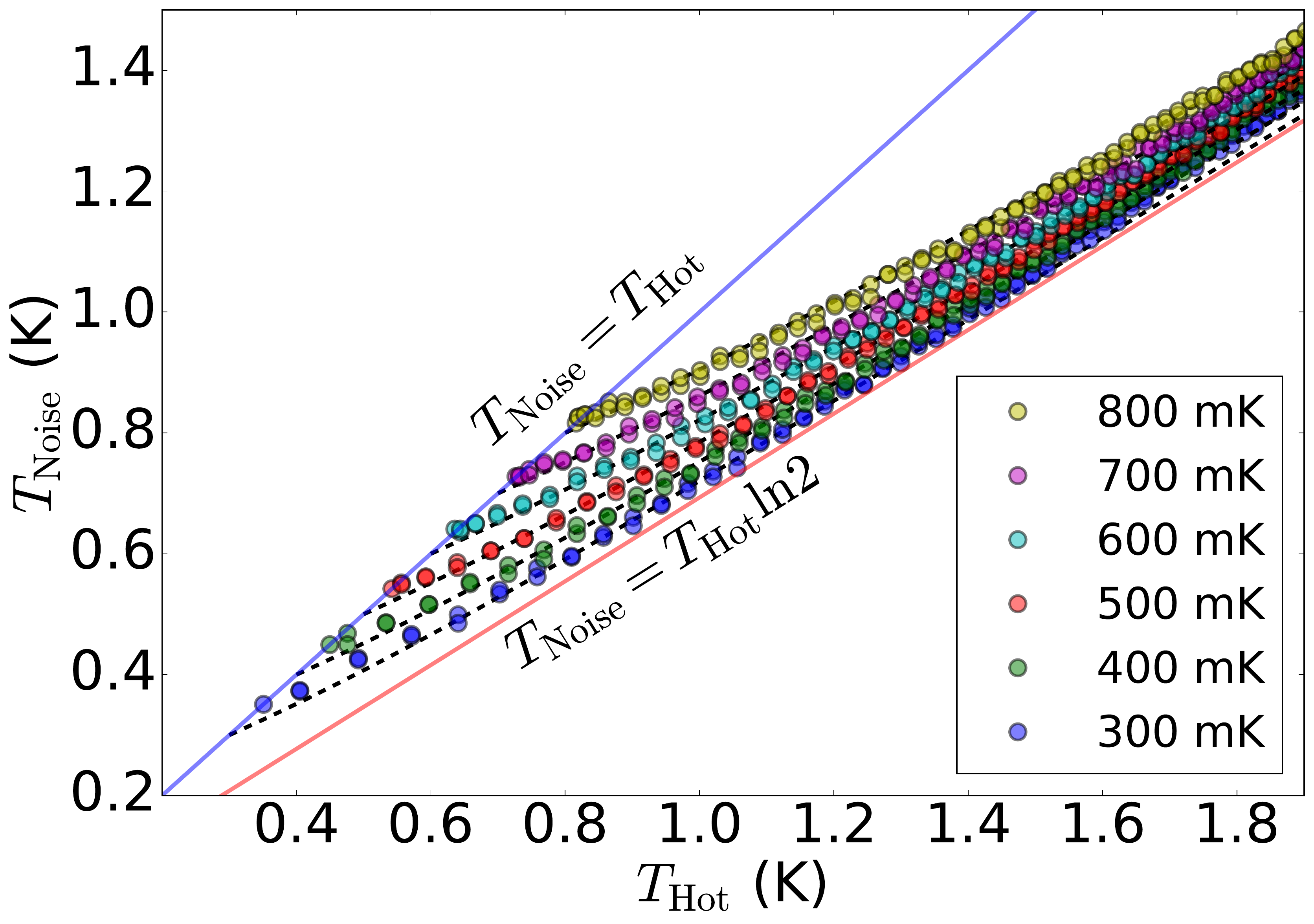}
	\caption{Noise temperature of the junction as a function of the temperature of the hot electrode. Symbols are experimental data taken at various phonon temperatures. The blue line represents the limit $T_{\text{Hot}} = T_{\text{Cold}}$ where the junction is at equilibrium. The red line represents the theory for $T_{\text{Cold}}=0$. Black dotted lines are numerical calculations for phonon temperatures from 300 mK to 800 mK.}
\label{fig:final}
\end{figure}

In order to be more quantitative and probe the validity of our analytical results of Eqs. (\ref{eq:TL=TR}) and (\ref{eq:TLggTR}), we now focus on the effect of a pure temperature difference with no voltage bias. We show in Fig. \ref{fig:final} the noise temperature of the junction as a function of the temperature of the hot electrode for various phonon temperatures (which is adjusted by heating the whole refrigerator). Colored circles are experimental data. The blue line corresponds to equilibrium, $T_{\text{Hot}}=T_{\text{Cold}}$ for which $T_{\text{Noise}}=T_{\text{Hot}}$. Data taken with no current in the wire fall on this line. For a small temperature difference $\Delta T=T_{\text{Hot}}-T_{\text{Cold}}$ we expect $T_{\text{Noise}}=T_{\text{Cold}}+\Delta T/2$. We indeed observe that experimental data leave the blue line with a slope close to $1/2$, as in \cite{lumbroso_electronic_2018}. The red line corresponds to the limit $T_{\text{Cold}}=0$ for which we expect $T_{\text{Noise}}=\mathrm{ln}\;2T_{\text{Hot}}$ according to Eq.(\ref{eq:TLggTR}). We clearly observe that experimental data at large $\Delta T$ tend to be parallel to the red line, showing the relevance of the $\mathrm{ln}\,2$ factor. Black dotted lines are theoretical predictions obtained numerically with no free parameter.  We observe an excellent agreement between theory and experiment for hot electrode temperatures up to 1.6 K. At higher temperature, experimental data start to deviate from the theory most probably because the too large power dissipated in the wire leads to an increase of the phonon temperature in the cold electrode. This causes a deviation of the data towards the Johnson-Nyquist limit (blue line) for which the two electrodes are at the same temperature.

\emph{Conclusion.}
We have measured the noise generated by a tunnel junction driven out-of-equilibrium by a temperature difference between its electrodes, i.e. the charge noise of a heat current flowing through the junction. Our results are in very good agreement with the Landauer-B\"{u}ttker theory of quantum transport. In particular we find that the noise temperature of the junction with one electrode at temperature $T$ and the other at much lower temperature is $T_{Noise}=T\text{ln}\;2$. This experiment paves the way to numerous new possibilities to explore noise and thermal properties of devices driven out-of-equilibrium by a thermal gradient. In the case of one electrode being near zero temperature, one expects the noise temperature of the device to be given by $T_{Noise}=F_TT$ with $F_T$ a "thermal Fano factor". In coherent samples like quantum point contacts, $F_T$ should be related to the usual charge Fano factor $F$. We find $F_T=1-F+2\text{ln}2F$, in agreement with the idea that the noise reveals the number of electron-holes excitation in the hot reservoir (the $\text{ln}2$ factor) through partitioning. However electron-electron interactions might affect $F$ and $F_T$ differently. This has been explored theoretically for diffusive wires in the hot electron regime \cite{sukhorukov_noise_2006}. Finally the use of fast noise detection techniques as developped in \cite{pinsolle_direct_2016} could be applied to our setup to detect the frequency dependence of the thermal conductivity of various samples, using an ac heated wire as an electrode whose temperature can be quickly modulated.

\begin{acknowledgments}
We acknowledge technical help of G. Lalibert\'{e}.
This work was supported by the Canada Excellence Research Chair program, the NSERC, the CFREF, the MDEIE, the FRQNT via the INTRIQ, the Universit\'{e} de Sherbrooke via the EPIQ and the Canada Foundation for Innovation.
\end{acknowledgments}

\bibliographystyle{apsrev4-1}
\bibliography{./jonction_tbias.bib}

\end{document}